\documentclass[aip, reprint]{revtex4-2}

\usepackage[utf8]{inputenc} 
\usepackage[T1]{fontenc}    
\usepackage{mathptmx}       
\usepackage{graphicx}       
\usepackage{dcolumn}        
\usepackage{bm}             
\usepackage{etoolbox}       
\usepackage{color}          

\makeatletter
\def\@email#1#2{%
 \endgroup
 \patchcmd{\titleblock@produce}
  {\frontmatter@RRAPformat}
  {\frontmatter@RRAPformat{\produce@RRAP{*#1\href{mailto:#2}{#2}}}\frontmatter@RRAPformat}
  {}{}
}%
\makeatother
\begin{document}


\title{Reproducible generation of green-emitting color centers in hBN using oxygen annealing} 

\author{\hspace{3cm} \\ 
Helmi Fartas$^{1,2}$, Sa\"id Hassani$^1$, Jean-Pierre Hermier$^1$, Ngoc Diep Lai$^2$, St\'ephanie Buil$^1$, Aymeric Delteil$^1$}
\affiliation{$^1$ Universit\'e Paris-Saclay, UVSQ, CNRS,  GEMaC, 78000, Versailles, France.\\$^2$ 
Université Paris-Saclay, ENS Paris-Saclay, CNRS, LuMIn, 91190, Gif-sur-Yvette, France.}



\begin{abstract}
\hspace{3cm}

The ability to generate quantum emitters with reproducible properties in solid-state matrices is crucial for quantum technologies. Here, we show that a high density of close-to-identical single-photon emitters can be created in commercial hexagonal boron nitride using annealing under oxygen atmosphere. This simple procedure yields a uniform in-plane distribution of color centers consistently emitting around 539.4~nm, with a wavelength spread smaller than 1~nm. We present an extensive characterization of their photophysical properties, showing that the emitters are bright and stable, and exhibit narrow lines at low temperatures with minimal spectral diffusion. These characteristics make this family of quantum emitters highly appealing for applications to quantum information science.

\hspace{3cm}
\end{abstract}


\maketitle 

Hexagonal boron nitride (hBN) has gained attention as an advantageous material platform for integrated quantum photonics~\cite{Wang19, Pelucci21}. As a van der Waals (vdW) material, hBN supports diverse fabrication techniques that enable its integration with other materials, including vdW crystals, in highly compact and complex devices, down to monolayer thicknesses. Moreover, hBN hosts optically active point defects that act as efficient single-photon emitters (SPEs) across various wavelength ranges~\cite{Tran16,Bourrelier16,Martinez16}. While some of these defects occur naturally, primarily near crystal edges or structural imperfections, several activation techniques, such as high-temperature annealing~\cite{Chen21} or surface etching~\cite{Chejanovsky16}, have been shown to yield a more uniform spatial distribution of the SPEs, away from flakes boundaries and defects. 

Nonetheless, the broad wavelength variability of hBN color centers poses a challenge for scalable integration into photonic devices. While most methods yield SPEs with strongly inhomogeneous properties -- and in particular a large wavelength spread --, a few methods have been recently identified to generate SPEs with reproducible wavelength using electron irradiation~\cite{Shevitski19, Fournier21, Gale22, Kumar23} and carbon implantation~\cite{Zhong24,Hua24}. Here, we introduce a simple and efficient method to reproducibly create a large number of SPEs through the annealing of commercial hBN under oxygen atmosphere. We generate and optically characterize SPEs emitting in the green wavelength range, with a narrow zero-phonon line (ZPL) and a minimal wavelength spread. The generated emitters are bright and stable, which is essential for quantum information applications.

 \begin{figure}
 \includegraphics[width=0.75\linewidth]{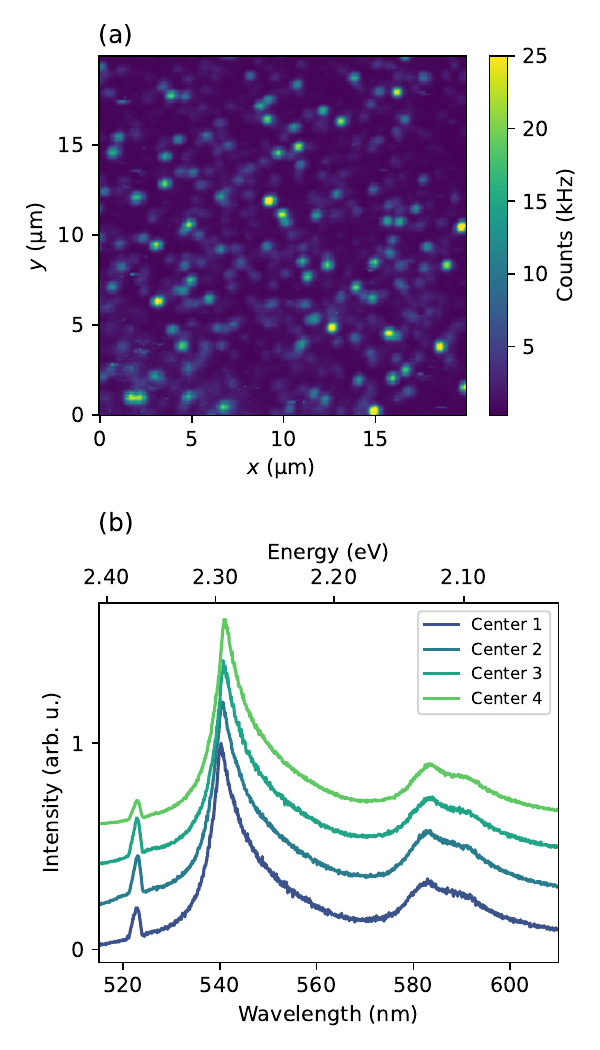}%
 \caption{\label{fig1} (a) Confocal map of a $12 \times 12$~$\mu$m$^2$ area of the flake. (b) Room-temperature photoluminescence spectra of 4 representative emitters. The 523~nm peaks originates from Raman scattering of the 485~nm laser.}%
 \end{figure}
  
 \begin{figure*}
 \includegraphics[width=\linewidth]{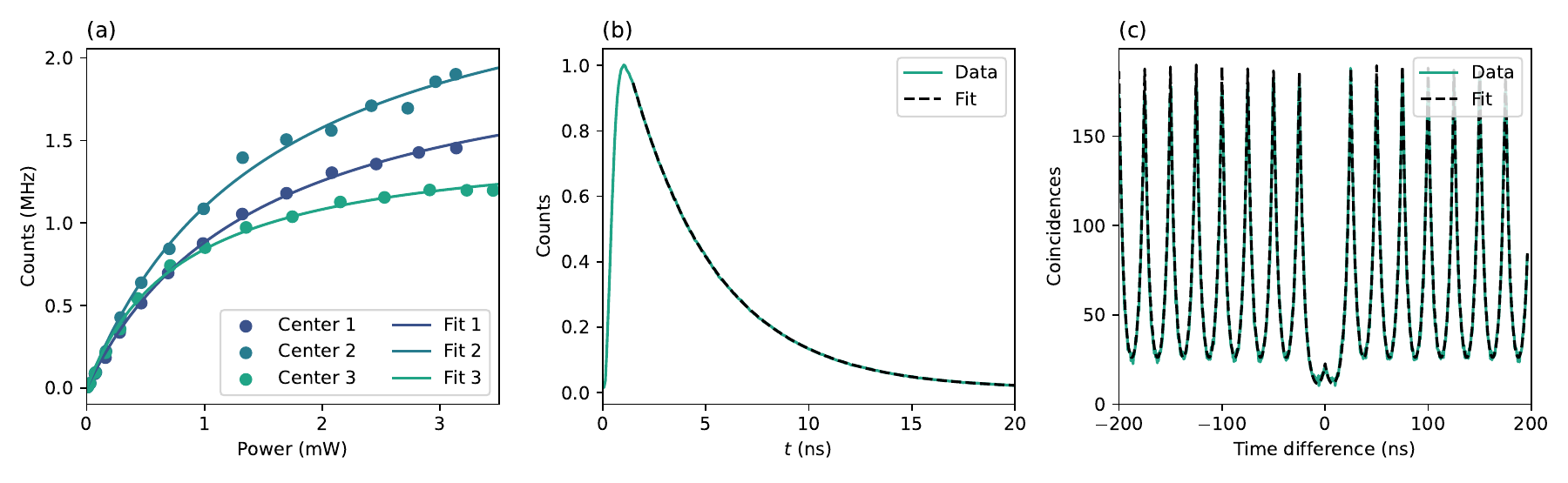}%
 \caption{\label{fig2} Room temperature photophysics of the green centers. (a) Count rate as a function of the laser power measured in CW regime for three emitters, revealing the saturation of the emitters. (b) Fluorescence decay in pulsed regime. The dashed line is a fit to the data. (c) Second-order photon correlations in pulsed regime, as a function of the time delay between detections. The dashed line is a fit to the data.}%
 \end{figure*}

 \begin{figure}
 \includegraphics[width=0.85\linewidth]{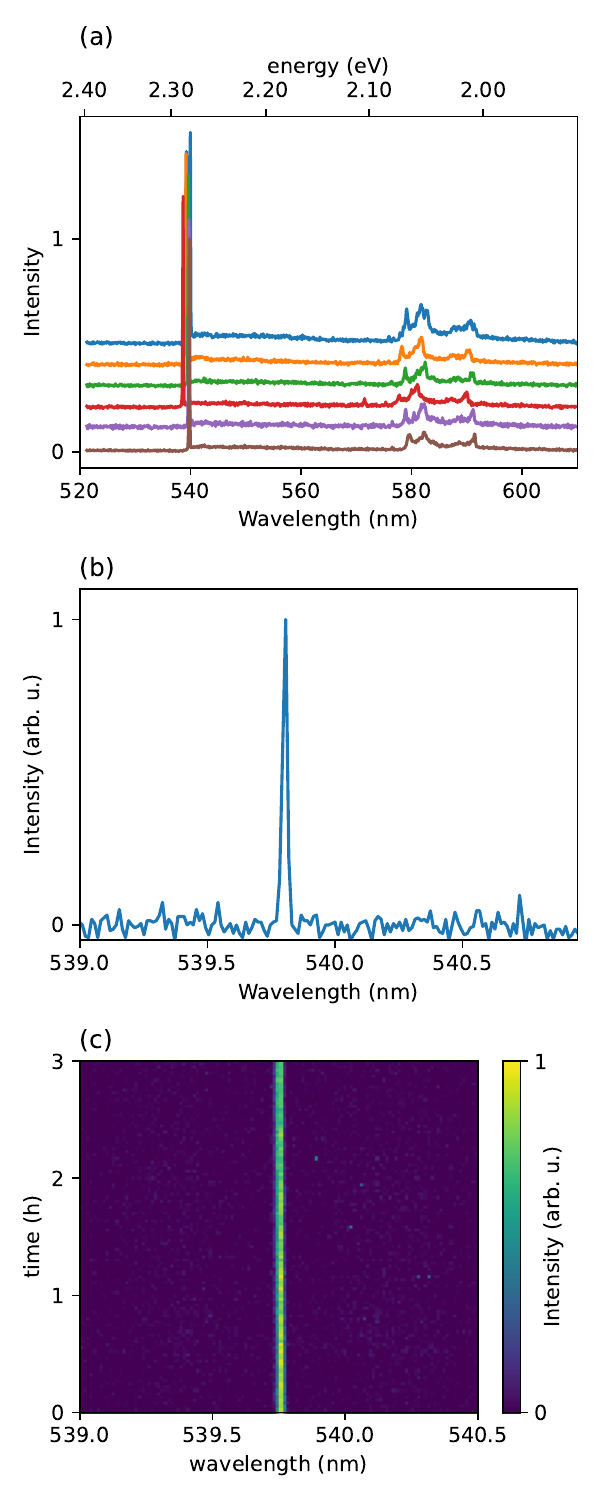}%
 \caption{\label{fig3} Low-temperature spectroscopy. (a) Low-resolution of the PL spectrum of 6~emitters. (b) High-resolution spectrum of the ZPL of a SPE. (c) Time-dependence of the ZPL emission as measured during 3~h. }%
 \end{figure}

The hBN crystals are exfoliated from commercial hBN (HQ Graphene) on SiO$_2$/Si substrate, which is subsequently annealed at 1000~$^{\circ}$C under 1000~sccm oxygen flow during 2~hours. The sample is then characterized in photoluminescence in a confocal microscope at room and cryogenic temperature. The sample is optically excited by a 485~nm diode laser which can perform both in continuous-wave (CW) and pulsed regimes. The laser is focused on the sample by an air objective of numerical aperture 0.95 located on top of the sample. The emitted photons are collected by the same objective, and channeled to a spectrometer or avalanche photodiodes with a resolution of 65~ps. We first performe confocal mapping by scanning the sample position with piezoelectric positioners. Fig.~\ref{fig1}a shows a typical confocal map measured at room temperature on a flake of thickness 156~nm, as measured by atomic force microscopy. The confocal map exhibits multiple localized spots, which correspond to emitters activated by the annealing process. Their repartition is quite uniform, and their density is high ($\sim 1~\mu$m$^{-2}$), yet not so high as to hinder individual addressing. Indeed, the vast majority of localized spots correspond to individual emitters. We then perform spectroscopy of individual emitters by characterizing the photon emission using a grating spectrometer. Fig.~\ref{fig1}b shows a few representative examples of emission spectra at room temperature. The emission exhibits a zero-phonon line (ZPL) at 540~nm, with a linewidth of about 9~nm. It displays a pronounced kurtosis, a characteristic already observed in other families of hBN emitters~\cite{Tran16ACS}, which takes its origin in emitter-phonon coupling~\cite{Wigger19}. Additionally, a bimodal optical phonon replica can be observed at 580 and 590~nm. This 165-200~meV replica is visible in most other quantum emitters in hBN~\cite{Tran16, Wigger19,Khatri19}.

We then perform characterizations aiming to assess the potential of the SPEs for room temperature operation. To infer the emitter brightness, we measure the count rate $C$ as a function of laser power $P$ in CW regime. As shown on Fig.~\ref{fig2}a, the SPEs exhibit a saturation behavior that can be fitted with the standard two-level saturation function $C = C_\infty / (1 + P_\mathrm{sat}/P)$, with $P_\mathrm{sat}$ the saturation power, found in the 1-1.5~mW range, and $C_\infty$ the asymptotic count rate. In many cases, $C_\infty$ is close to, or higher than, 2~million counts per second. Given that this count rate is measured out of any photonic structure, and most of the emission is directed toward the high index substrate, this places the emitter among the brightest emitters in this material platform. Then, the emitter lifetime is measured in pulsed regime. Fig.~\ref{fig2}b shows an example of photoluminescence decay. It can be fitted with a single-exponential function providing $\tau = 4.2$~ns. This decay time lies in the typical range for hBN color centers. The combined measurement of saturation count rate $C_\infty$ and emitter lifetime leads us to expect a high quantum efficiency for this family of emitters. A confirmation would require the observation of Purcell modification of the decay time~\cite{Gerard24}. Another important figure of merit is the photon purity, which can be assessed by measuring the second-order correlation function in a Hanbury Brown and Twiss setup. Fig.~\ref{fig2}c shows the measured $g^{(2)}(\tau)$ as a function of the time delay $\tau$ between photon detections, measured in pulsed regime during 3~min. It allows us to extract $g^{(2)}(0) = 0.09 \pm 0.01$, which unequivocally demonstrates single-photon emission.
 
To further investigate the potential of the green emitters for experiments based on Hong-Ou-Mandel interference, we perform additional characterization at cryogenic temperature, where the emission spectrum narrows down drastically. Fig.~\ref{fig3}a shows eight SPE spectra measured at 4~K. All exhibit a narrow ZPL, as well as phonon replica with a complex substructure, which could help identifying the defect from ab-initio calculations. We note that the lineshape of the phonon replica is similar to that of ref~\cite{Khatri19}. Although the emission wavelength is different, this suggests a comparable emitter-phonon coupling. Fig.~\ref{fig3}b shows a higher resolution spectrum of the ZPL, yet the observed linewidth is still limited by the spectrometer resolution ($120~\mu$eV). Inferring the exact linewidth will require more advanced techniques, such as resonant excitation~\cite{Horder22, Fournier23, Gerard25} or interferometry~\cite{Brokmann06,Spokoyny20}. Fig.~\ref{fig3}c shows the evolution of the emission spectrum as a function of time during 3~hours. No spectral fluctuations can be observed at our spectral resolution. Additionally, we notice no intensity fluctuations or blinking, demonstrating the remarkable stability of the SPE. 

\begin{figure}[h]
 \includegraphics[width=0.8\linewidth]{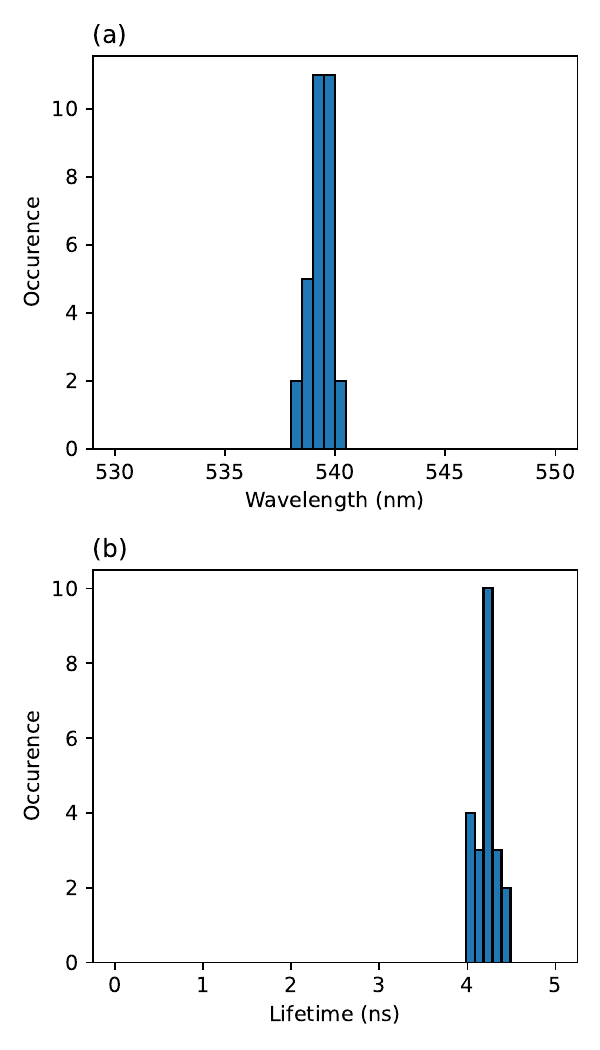}%
 \caption{\label{fig4} (a) Histogram of the ZPL center wavelength. (b) Histogram of the emitter lifetime.}%
 \end{figure}

To obtain further insight into the statistical dispersion of the SPE properties, on Fig.~\ref{fig4}a we show a histogram of the ZPL position for 30~emitters. The distribution is centered at 539.35~nm, and has a standard deviation of 0.45~nm. This ensemble distribution is remarkably narrow for hBN emitters, only comparable to that of B-centers~\cite{Fournier21}. We have also performed a statistical analysis of the SPE lifetime, shown on Fig.~\ref{fig4}b, which yields $\tau = 4.22 \pm 0.10$~ns, showing the high reproducibility of the emitter photophysical properties.

\begin{figure}[h]
 \includegraphics[width=0.85\linewidth]{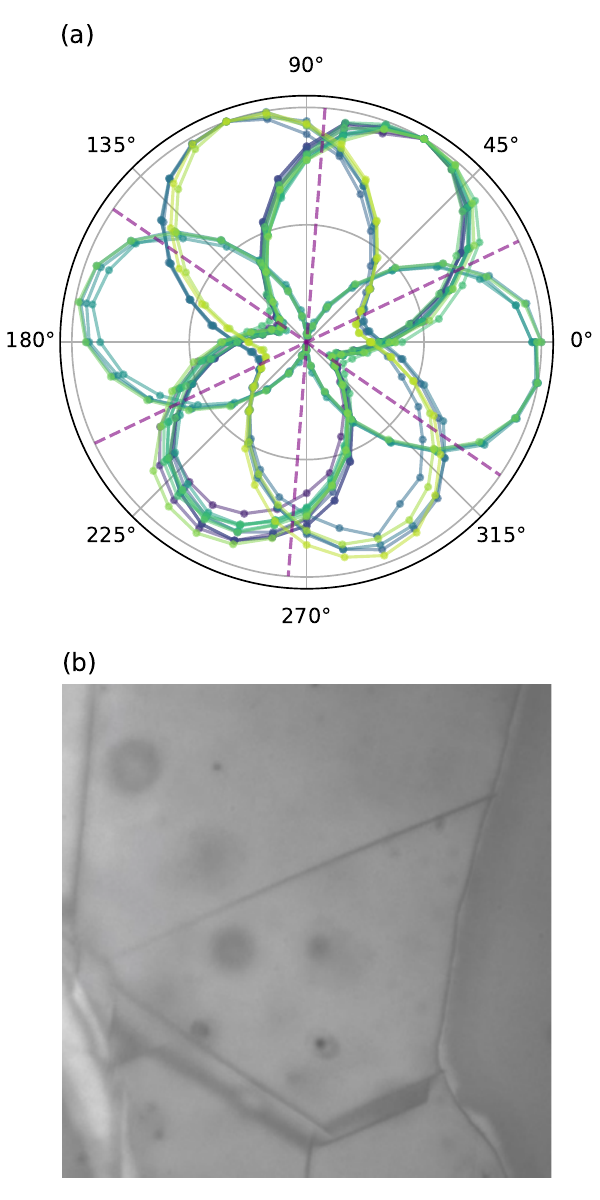}%
 \caption{\label{fig5} (a) Count rate as a function of the polarizer angle measured for 20~emitters. The dashed lines represent the crystal axes visible on Fig.~\ref{fig5}b. (b) Optical microscope image of the flake, revealing the crystal axes.}%
 \end{figure}

We characterize the orientation of the emission dipole at room temperature by inserting a polarizer on the detection path and measuring the count rate as a function of the polarizer angle. The excitation laser polarization is set to a circular polarization. Fig.~\ref{fig5}a shows the emission polarization of a few representative SPEs. All are linearly polarized, indicating an in-plane dipole -- like the vast majority of hBN SPEs. The polarization axes are bunched in three groups, as is the case of SPEs created by carbon implantation~\cite{Zhong24} and electron irradiation~\cite{Horder24}. This is consistent with a $C_{2v}$ symmetry of the defect center. The polarization axes are close to crystal axes as identified by the twin boundaries~\cite{Rooney18} visible on optical and SEM images (Fig.~\ref{fig5}b). However, further investigations are needed to establish the type of crystal direction (\textit{i.e.} zigzag or armchair) along which the dipole is oriented, which should provide helpful information for defect identification.

While the nature of the green SPEs is unknown, oxygen is likely a constituent of the color center, since we verified that annealing under the same conditions but without oxygen does not generate any emitter. Moreover, the defect formation also possibly involves a preexisting impurity in the crystal. Indeed, while we verified that the green defects can be reproducibly generated in commercial hBN (HQ Graphene), on the other hand we could not generate them in high-purity hBN grown by the high-pressure, high-temperature technique~\cite{Taniguchi07}. Instead, oxygen annealing of the latter material yields quantum emitters in the red to infrared spectral range, with no wavelength control~\cite{Chen21}. We also note that it is possible to generate color centers emitting in the infrared range using a different procedure~\cite{Mohajerani24}, showing the variety and complexity of quantum emitter generation in hBN. Oxygen is expected to form stable defects in hBN, both in substitutional positions and in complexes with vacancies.~\cite{Tawfik17, Li22, Na21}. Additionally, it can also form complexes with preexisting impurities such as carbon~\cite{Cholsuk24} -- commonly found natively in hBN crystals depending on the growth process. Finally, we performed ODMR experiments, without detecting any signal in the range 0-8~GHz, both with and without external magnetic field, which discards easily accessible spin-dependent optical transitions, in contrast to other families of hBN defects~\cite{Gottscholl20}. Further investigations will be needed to unequivocally identify the defect at the origin of the green emission, based on additional characterization combined with DFT calculations~\cite{Tawfik17}.

To conclude, we presented a general investigation of the photophysics of a new family of quantum emitters in hBN that can be reproducibly generated at a large scale, and that exhibits remarkable photophysical properties, with, in particular, a highly reproducible emission wavelength and lifetime. Its narrow linewidth and high brightness makes it an appealing contender for applications to quantum photonics and quantum information science. Our study also provides valuable insights for future identification of this color center. \\
~

\begin{acknowledgments}

We acknowledge D. G\'erard, C. Arnold and T. Berg\`ese for fruitful discussions and technical assistance. This work is supported by the French Agence Nationale de la Recherche (ANR) under reference ANR-21-CE47-0004-01. This work has also been supported by Region \^Ile-de-France in the framework of DIM QuanTiP.
\end{acknowledgments}

~\\
%
%

\end{document}